\documentclass[aps,prb,superscriptaddress,reprint,floatfix,showpacs]{revtex4-1}
\usepackage{graphicx}
\usepackage{dcolumn}
\usepackage{bm}
\usepackage{natbib}
\usepackage{longtable}
\usepackage{amsmath}
\usepackage{url}

\begin{document}

\title{Orbital magnetic moment and coercivity of SiO$_{2}$-coated FePt nanoparticles studied by x-ray magnetic circular dichroism}

\author{Y. Takahashi}
\email{takahashi@wyvern.phys.s.u-tokyo.ac.jp}
\affiliation{Department of Physics, University of Tokyo, Bunkyo-ku, Tokyo 113-0033, Japan}
\author{T. Kadono}
\affiliation{Department of Physics, University of Tokyo, Bunkyo-ku, Tokyo 113-0033, Japan}
\author{V. R. Singh}
\affiliation{Department of Physics, University of Tokyo, Bunkyo-ku, Tokyo 113-0033, Japan}
\author{V. K. Verma}
\affiliation{Department of Physics, University of Tokyo, Bunkyo-ku, Tokyo 113-0033, Japan}
\author{K. Ishigami}
\affiliation{Graduate School of Frontier Science, University of Tokyo, Bunkyo-ku, Tokyo 113-0033, Japan}
\author{G. Shibata}
\affiliation{Department of Physics, University of Tokyo, Bunkyo-ku, Tokyo 113-0033, Japan}
\author{T. Harano}
\affiliation{Department of Physics, University of Tokyo, Bunkyo-ku, Tokyo 113-0033, Japan}
\author{A. Fujimori}
\affiliation{Department of Physics, University of Tokyo, Bunkyo-ku, Tokyo 113-0033, Japan}
\affiliation{Quantum Beam Science Directorate, Japan Atomic Energy Agency, Sayo-cho, Sayo-gun, Hyogo 679-5148, Japan}
\author{Y. Takeda}
\affiliation{Quantum Beam Science Directorate, Japan Atomic Energy Agency, Sayo-cho, Sayo-gun, Hyogo 679-5148, Japan}
\author{T. Okane}
\affiliation{Quantum Beam Science Directorate, Japan Atomic Energy Agency, Sayo-cho, Sayo-gun, Hyogo 679-5148, Japan}
\author{Y. Saitoh}
\affiliation{Quantum Beam Science Directorate, Japan Atomic Energy Agency, Sayo-cho, Sayo-gun, Hyogo 679-5148, Japan}
\author{H. Yamagami}
\affiliation{Quantum Beam Science Directorate, Japan Atomic Energy Agency, Sayo-cho, Sayo-gun, Hyogo 679-5148, Japan}
\affiliation{Department of Physics, Kyoto Sangyo University, Motoyama, Kamigamo, Kita-Ku, Kyoto 603-8555, Japan}

\author{S. Yamamoto}
\affiliation{Institute for Integrated Cell-Material Science (iCeMS), Kyoto University, Yoshida Ushinomiya-cho, Sakyo-ku, Kyoto 606-8501, Japan}
\author{M. Takano}
\affiliation{Institute for Integrated Cell-Material Science (iCeMS), Kyoto University, Yoshida Ushinomiya-cho, Sakyo-ku, Kyoto 606-8501, Japan}

\begin{abstract}
We have investigated the spin and orbital magnetic moments of Fe in FePt
 nanoparticles in the $L$1$_{0}$-ordered phase coated with SiO$_{2}$ by
 x-ray absorption spectroscopy (XAS) and x-ray magnetic circular
 dichroism (XMCD) measurements at the Fe $L_{\rm 2,3}$ absorption
 edges. Using XMCD sum rules, we evaluated the ratio of the orbital
 magnetic moment ($M_{\rm orb}$) to the spin magnetic moment ($M_{\rm
 spin}$) of Fe to be $M_{\rm orb}/M_{\rm spin}$ = 0.08. This $M_{\rm
 orb}/M_{\rm spin}$ value is comparable to the value (0.09) obtained for FePt
 nanoparticles prepared by gas phase condensation, and is larger than the
 values ($\sim$0.05) obtained for
 FePt thin films, indicating a high
 degree of $L$1$_{0}$ order. The hysteretic behavior of the FePt
 component of the magnetization was measured by XMCD. The magnetic coercivity ($H_{\rm c}$) was found to
 be as large as 1.8 T at room temperature, $\sim$3 times larger than the
 thin film value and $\sim$50 times larger than that of the gas phase
 condensed nanoparticles. The hysteresis curve is well explained by the
 Stoner-Wohlfarth model for non-interacting single-domain nanoparticles
 with the $H_{\rm c}$ distributed from 1 T to 5 T.
\end{abstract}


\maketitle

\indent The $L$1$_{0}$-ordered alloy FePt has attracted great attention as
a material to be used in high-density storage memory devices because
FePt possesses a high magneto-crystalline anisotropy in bulk FePt and
thin films
\cite{Ivanov197315,Okamoto2002Chemical-order-,Seki2006Spin-polarized-,Seki2008Nucleation-type}. In
order to increase the storage density of memory devices, FePt
nanoparticles are promising materials. FePt nanoparticles prepared by
the gas phase preparation method \cite{Dmitrieva2006Enhancement-of-} and
the wet chemical preparation method \cite{Yamamoto2006Preparation-of-},
however, have the fcc structure, in which magneto-crystalline anisotropy
disappears. By annealing the nanoparticles at high temperatures, one can
transform the structure from the fcc to $L$1$_{0}$, but then the
nanoparticles are easily oxidized and are covered by surface oxides. The plasma treatments of FePt nanoparticles followed by {\it in situ} flight
annealing have been performed to remove
the oxidized surfaces \cite{Boyen2005Electronic-and-,Dmitrieva2007Magnetic-moment}, however, it has simultaneously reduced the magnetic coercivity
($H_{\rm c}$) as small as 0.038 T \cite{Dmitrieva2007Magnetic-moment}. In addition, high temperature annealing is difficult to implement in
existing manufacturing process because it may lead to the coalescence of
nanoparticles into larger particles
\cite{Yamamoto2005Magnetically-su}. Recently, in order to overcome such
difficulties, FePt nanoparticles coated with SiO$_{2}$ were prepared and
dispersed before annealing to achieve a high degree of $L$1$_{0}$ order
\cite{Morimoto200615,Tamada2006Microscopic-Cha,Tamada2007Well-ordered-L1,Tamada200715,Yamamoto2005Magnetically-su,Yamamoto2006Preparation-of-}. FePt
nanoparticles prepared by this method have a $H_{\rm c}$ as
large as 1.85 T at 300 K \cite{Yamamoto2005Magnetically-su}, of the same
order as the value of 3.0 T
observed for
$L$1$_{0}$-FePt (001) particulate films \cite{Wang2009Phenomenologica}, and
much larger than the bulk value of 0.16 T \cite{Tanaka199715}, the thin
film value of 0.55 T \cite{Seki2008Nucleation-type}, and the value of 0.75 T
for microfabricated FePt (001) dots \cite{Wang2008Magnetization-r}.\\
\indent In order to understand the origin of the large $H_{\rm c}$ and
to further improve the properties of the SiO$_{2}$-coated FePt
nanoparticles, microscopic information about the magnetism such as the
orbital magnetic moment is important. The chemical state of Fe can be easily distinguished by soft x-ray absorption spectroscopy (XAS), and the orbital and spin magnetic moments of
transition-metal 3$d$ orbitals can be measured by soft x-ray magnetic
circular dichroism (XMCD) measurements of the 2$p$ core level of the
transition element \cite{Thole1992x-ray-circular-}. Furthermore, in XMCD
measurements, one can eliminate extrinsic magnetic signals such as those
from oxidized Fe and those from the
diamagnetic SiO$_{2}$ coating. In the present work, we have applied the
XAS and XMCD techniques to the SiO$_{2}$-coated FePt nanoparticles to characterize
their magnetic properties from the microscopic level.\\
\indent FePt nanoparticle samples were prepared by the wet chemical
method. Precursor fcc FePt nanoparticles were prepared according to the
method of Sun {\it et al.} \cite{Sun2000Monodisperse-Fe}, and were
subsequently coated by SiO$_{2}$ according to the method of Fan {\it et
al.} \cite{Fan2004Self-Assembly-o}. Thus during the heat treatment
thermal diffusion of Fe and Pt atoms was confined inside the SiO$_{2}$
nanoreactor. For details of the fabrication process, we refer the reader
to Yamamoto {\it et al.} \cite{Yamamoto2005Magnetically-su}. For the
XMCD measurements, the nanoparticle samples were supported by silver
paste on a sample holder. After the preparation of the samples, they
were kept in an Ar atmosphere in order to prevent oxidation until the
XMCD measurements. XAS [=($\mu_{+}+\mu_{-}$)/2] and
XMCD (=$\mu_{+}-\mu_{-}$) spectra were obtained at the helical
undulator beam line BL23SU of SPring-8. Here, $\mu_{+}$ and $\mu_{-}$
are XAS spectra taken with the light helicity parallel and antiparallel
to the incident photon direction, respectively. The highest applied
magnetic field was 9 T. The measurements were done under an ultra-high
vacuum of $\sim 4.8\times 10^{-9}$ Pa. The polarity of the synchrotron
radiation was switched at each photon energy using a kicker magnet with
the frequency of 1 Hz \cite{Saitoh2012Performance-upg}. The measurements were done at room
temperature ($\sim$300 K) in the total electron yield (TEY) mode. The
magnetic field dependence of the XMCD intensity at the peak of the Fe $L_{\rm 3}$
edge was measured in the range of $-9\ {\rm T}\leq H\leq 9\ {\rm T}$.\\ 
 \begin{figure}
 \begin{center}
\includegraphics[width=86mm]{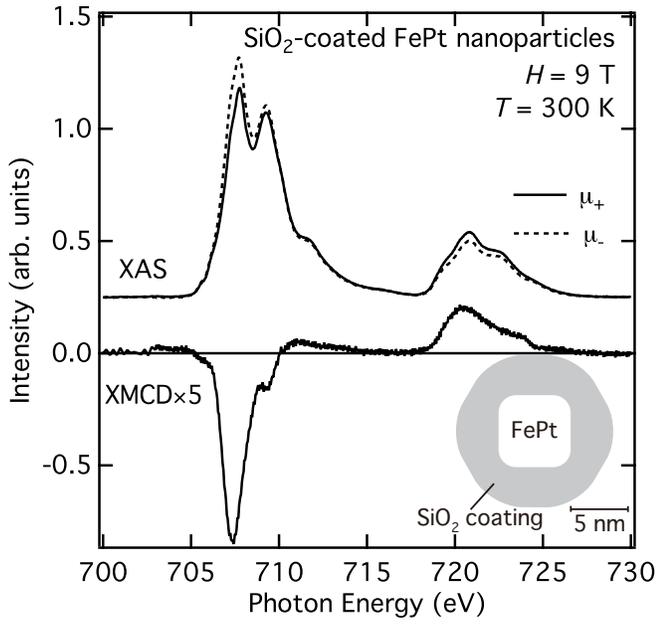}
\caption{\label{xasxmcd}(Color online) XAS and XMCD spectra at the Fe $L_{\rm 2,3}$
  edges of the SiO$_{2}$-coated FePt nanoparticles under the applied magnetic field of 9 T. All the spectra have been
  normalized to the $L_{\rm 3}$ XAS peak intensity. Inset shows a schematic figure of
  the FePt nanoparticle coated by SiO$_{2}$. According to transmission
  electron microscopy (TEM) pictures reported by Yamamoto {\it et al.}
  \cite{Yamamoto2005Magnetically-su}, the average size of the FePt
  nanoparticles is estimated to be 6-7 nm and that of the SiO$_{2}$ coat to be 10-15 nm.}
 \end{center}
 \end{figure}
\indent Figure \ref{xasxmcd} shows the XAS and XMCD spectra at the Fe $L_{\rm 2,3}$ edges of
the FePt nanoparticles. All the spectra have been normalized to the
$L_{\rm 3}$ XAS peak intensity. Unlike the Fe $L_{\rm 2,3}$ spectra of FePt by Boyen {\it et al.}
\cite{Boyen2005Electronic-and-} and Dmitrieva {\it et al.} \cite{Dmitrieva2007Magnetic-moment}, each of the $L_{\rm 3}$ and
$L_{\rm 2}$ edges exhibits a doublet structure: The $L_{\rm 3}$ doublet
peaks are located at 707.7 eV and 709.3 eV, and the $L_{\rm 2}$ ones at
720.8 eV and 722.7 eV. Furthermore, broad shoulder/tail structures are
observed on the high-energy side of the $L_{\rm 3}$ peak from 710 eV to
717 eV and on the high-energy side of the $L_{\rm 2}$ peak from 722 eV
to 730 eV. We consider that the origin of the doublets is overlapping
signals of FePt (at 707.7 and 720.8 eV) and Fe oxides (at 709.3 and
722.7 eV). The oxides are most likely extrinsic Fe$^{3+}$ oxides formed at the surface
of bare FePt particles whose SiO$_{2}$-coating was removed when grinding
the sample into fine powders, as one can see from the TEM picture [shown
in Fig. 2(b) of Yamamoto {\it et al.}
\cite{Yamamoto2005Magnetically-su}]. We tentatively attribute the
shoulder/tail structures to the charge-transfer 2$p^{5}d^{6}$-$2p^{5}d^{7}\underline{L}$
satellite of the Fe$^{3+}$ oxide \cite{Crocombette1995x-ray-absorptio} but the exact origin of the structure remains to be clarified. On the other hand, the lineshape of the experimental XMCD
spectrum in Fig. \ref{xasxmcd} is almost identical to that of FePt by Dmitrieva {\it et al.} \cite{Dmitrieva2007Magnetic-moment} except
for a small negative peak around 709 eV, indicating that contributions of the Fe
oxides to the XMCD spectrum are small.\\
 \begin{figure}
 \begin{center}
\includegraphics[width=86mm]{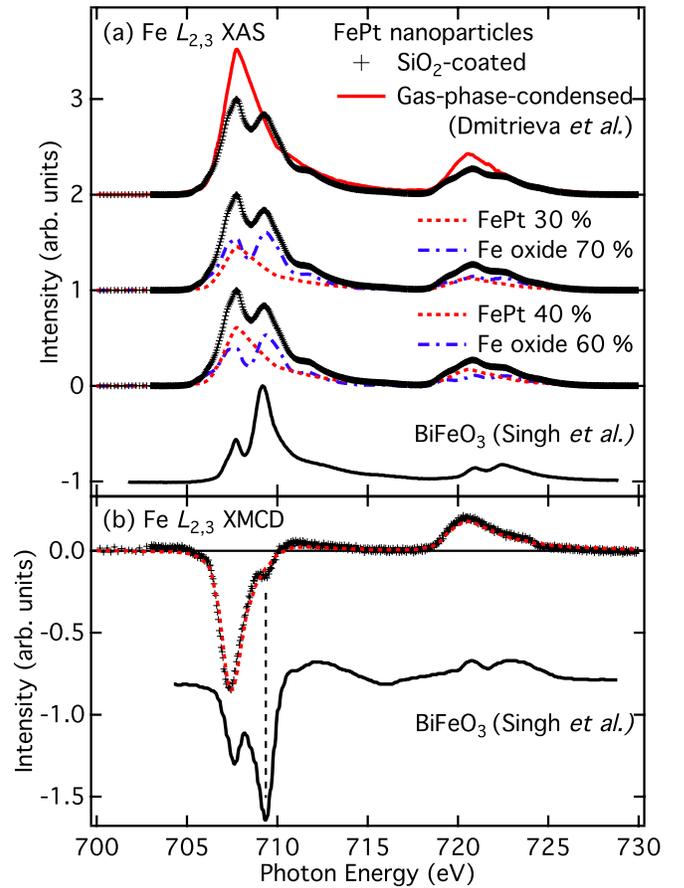}
\caption{\label{xas}(Color online) Decomposition of the measured XAS and XMCD spectra into those
  of FePt and a Fe$^{3+}$ oxide. (a) The experimental XAS spectrum and that of
  gas-phase-condensed FePt nanoparticles
  (Ref. \!\onlinecite{Dmitrieva2007Magnetic-moment}) are shown by
  crosses and a solid curve, respectively, at the top. Spectra
  after subtraction of the FePt ones (dotted curves) from the measured
  ones by different ratios are indicated below by dot-dashed curves. The XAS spectrum of
  BiFeO$_{3}$ by Singh {\it et al.}
  \cite{Singh2013Enhanced-ferrom} is shown at the bottom. (b) The experimental XMCD spectrum
  (crosses) and the
  FePt one (Ref. \!\onlinecite{Dmitrieva2007Magnetic-moment}, dotted curve) are
  shown at the top. The XMCD spectrum of BiFeO$_{3}$ by Singh {\it et al.}
  \cite{Singh2013Enhanced-ferrom} is shown at the bottom.}
 \end{center}
 \end{figure}
\indent In order to estimate signals of the Fe oxides, we first subtracted the FePt XAS spectrum reported by Dmitrieva {\it et
al.} \cite{Dmitrieva2007Magnetic-moment}, where Fe oxide was completely removed by plasma treatment, from the
experimental one. In Fig. \ref{xas}(a), one can see that after having
subtracted the FePt XAS spectrum multiplied by $\sim$0.4, the spectral lineshape became similar to
that of BiFeO$_{3}$, where the Fe$^{3+}$ ion is located at the $O_{\rm
h}$ site \cite{Singh2013Enhanced-ferrom},
shown in the bottom of Fig. \ref{xas}(a).
Thus the intrinsic signals from FePt are estimated to be 40 \% of the
total Fe $L_{\rm 2,3}$ XAS intensity. The experimental XMCD spectrum at
 the Fe $L_{\rm 2,3}$ edges of the FePt nanoparticles is shown in Fig. \ref{xas}(b). The lineshape of the XMCD
 spectrum of FePt nanoparticles prepared by the gas condensation method by Dmitrieva {\it et al.} \cite{Dmitrieva2007Magnetic-moment}
 is also shown by a dotted curve. Unlike the XAS spectra, both XMCD data agree with each other except for the small negative
 peak around 709 eV in the present sample. From comparison of the
 present XMCD spectrum with the XMCD spectrum of BiFeO$_{3}$ shown at the
 bottom of Fig. \ref{xas}(b), the small negative peak can be attributed to a peak of a Fe$^{3+}$ oxide. After having removed the Fe-oxide contributions from the experimental
 XAS and XMCD spectra, we applied
 the XMCD sum rules \cite{Carra1993x-ray-circular-,Thole1992x-ray-circular-}
  and obtained the spin moment $M_{\rm spin}$ = 1.4 $\mu_{\rm B}/{\rm
  Fe}$, the orbital moment $M_{\rm
 orb}$ = 0.12 $\mu_{\rm B}/{\rm Fe}$, and their ratio $M_{\rm
 orb}$/$M_{\rm spin}$ = 0.08. These values are
 summarized in Table \ref{table}, and are compared with those of different
 kinds of FePt samples \cite{Dmitrieva2007Magnetic-moment,Imada2007Perpendicular-m}. The obtained $M_{\rm
 orb}$/$M_{\rm spin}$ ratio is comparable to the value (0.09) obtained for FePt
 nanoparticles prepared by gas phase condensation \cite{Dmitrieva2007Magnetic-moment}, and is larger than the
 values ($\sim$0.05) obtained for
 FePt thin films \cite{Imada2007Perpendicular-m}, indicating a high
 degree of $L$1$_{0}$ order. The deduced $M_{\rm spin}=1.4\ \mu_{\rm
 B}/{\rm Fe}$ is, however, small compared to the other FePt samples
 ($\geq 2\ \mu_{\rm
 B}/{\rm Fe}$). The origin of this discrepancy is not clear at present,
 but the outermost layers of the bare FePt particles might also be
 oxidized to Fe$^{2+}$ and become non-ferromagnetic because the XAS and
 XMCD lineshapes of Fe$^{2+}$ oxides and metallic Fe are hardly distinguishable.  \\
\begin{table}
\caption{\label{table} Values of spin ($M_{\rm spin}$) and orbital
 ($M_{\rm orb}$) magnetic moments (in units of $\mu_{\rm
 B}/{\rm Fe}$) and the coercive field $H_{\rm c}$ (in units of T) for different FePt samples.}
\begin{tabular}{lcccc}\hline
&$M_{\rm spin}$&$M_{\rm orb}$&$M_{\rm orb}/M_{\rm spin}$&$H_{\rm c}$\\ \hline
Nanoparticle (present work)&1.4&0.12&0.08&1.8\\
Nanoparticle (gas phase)
 \cite{Dmitrieva2007Magnetic-moment}&2.21&0.19&0.09&0.038\\ 
FePt thin films \cite{Imada2007Perpendicular-m}&0.5&0.025&0.05$\pm$0.01&0.1\\ \hline
\end{tabular}
\end{table}
\begin{figure}
 \begin{center}
\includegraphics[width=86mm]{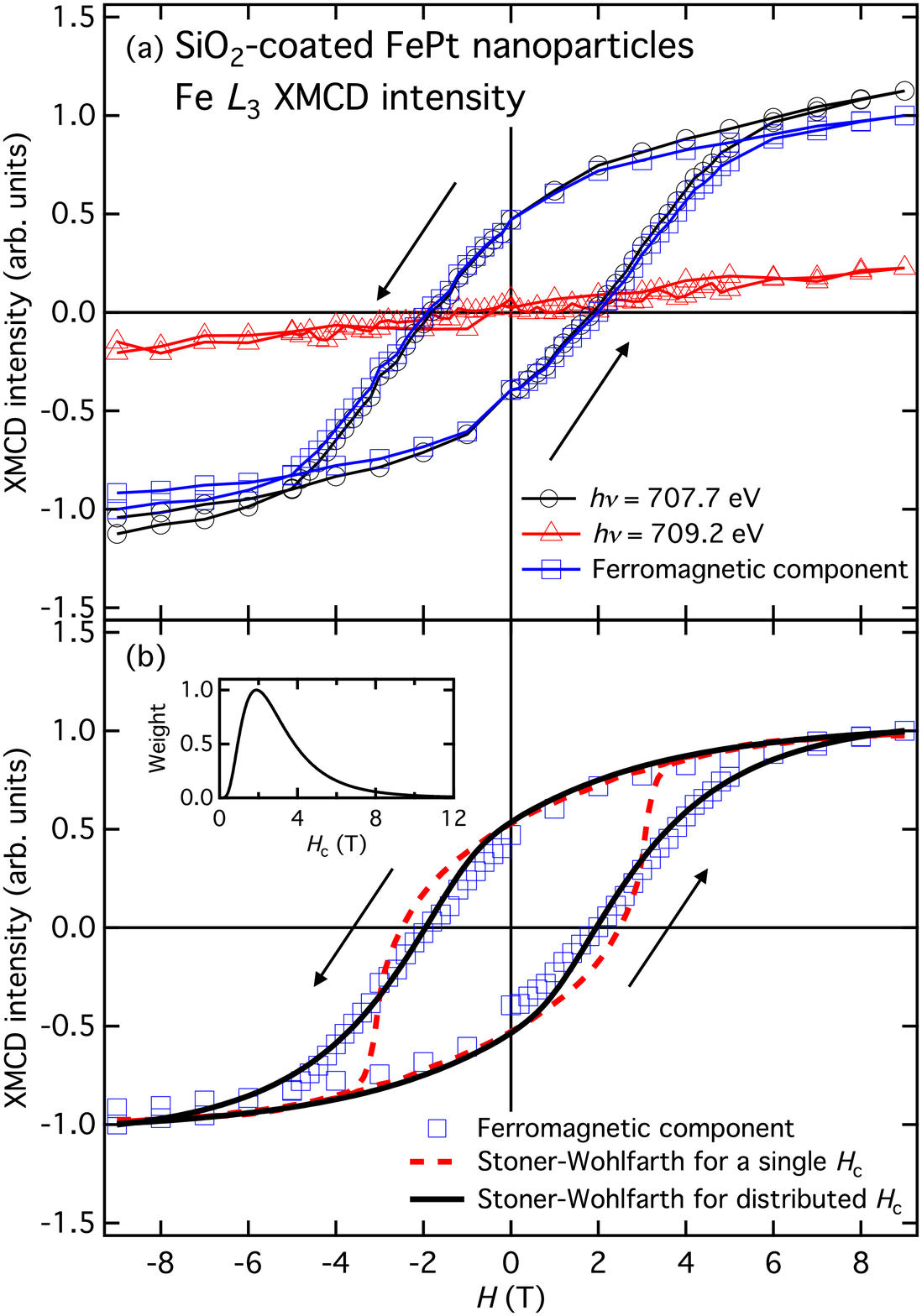}
\caption{\label{hys}(Color online) Hysteresis loops of the XMCD intensities (normalized
  to the saturation magnetization) at the Fe
  $L_{\rm 3}$ edge of the SiO$_{2}$-coated FePt nanoparticles. (a)
  Hysteresis loops at $h\nu$ = 707.7 eV (open circles) and 709.2 eV
  (open triangles)
  corresponding to a peak due to FePt and the Fe$^{3+}$ oxides,
  respectively. A ferromagnetic component extracted from the hysteresis
  loop at $h\nu$ = 707.7 eV (open squares) is also plotted. (b)
  The ferromagnetic component of the XMCD hysteresis curve compared with a
  calculated Stoner-Wohlfarth hysteresis loop
  \cite{Coey2010Magnetism-and-M} for a single $H_{\rm c}$
  (= 2.5 T) (dashed curve) and for distributed $H_{\rm c}$ (solid
  curve). Inset shows the log-normal
  distribution of $H_{\rm c}$.}
 \end{center}
 \end{figure}
\indent Figure \ref{hys}(a) shows the Fe $L_{\rm 3}$-edge XMCD
intensities of the FePt nanoparticles as functions of
magnetic field at two different photon energies. The curve at
$h\nu$ = 709.2 eV, which reflects the XMCD signal of the Fe$^{3+}$ oxide,  is linear in $H$ and shows no hysteretic behavior. Since
the slope is given by the magnetic susceptibility $\chi$ of the Fe$^{3+}$ ion of
the oxide, it is given by $C/T$ if the Fe$^{3+}$ ion is
paramagnetic or by $C/T_{\rm N}$ if it is antiferromagnetic. Here, $C$
and $T_{\rm N}$ are the Curie constant for the Fe$^{3+}$ ($S=5/2$) ion
and the Neel temperature, respectively. From the slope $\chi\sim 0.023\
\mu_{\rm B}/{\rm T}$, we estimated $T$ or $T_{\rm N}\sim C/\chi =325$ K,
indicating that the Fe$^{3+}$ at 300 K are paramagnetic or slightly below
$T_{\rm N}$, and cannot be ferro/ferrimagnetic.\\
\indent In contrast, the XMCD intensity at
$h\nu$ = 707.7 eV, which is dominated by FePt, shows a hysteretic
behavior with $H_{\rm c}$ of 1.8 T. Because a $M$-$H$ curve of the peak
of the Fe oxides overlaps with the curve at $h\nu$ = 707.7 eV, we
obtained a $M$-$H$ curve of a ferromagnetic component by subtracting the
curve at $h\nu$ = 709.2 eV from the curve at $h\nu$ = 707.7 eV. From the shape of the hysteresis
loop thus obtained as shown
in Fig. \ref{hys}(a), one can see that the value of the remnant magnetization is half of
that of the saturation
magnetization and the shape itself is somewhat rounded compared to a
typical rectangular hysteresis loop for ferromagnetic materials. These
behaviors are
characteristic of the hysteresis of the Stoner-Wohlfarth model
\cite{Stoner1948A-Mechanism-of-,Coey2010Magnetism-and-M}, which explains $H_{\rm c}$ based on the coherent
reversal in non-interacting single-domain particles and in good
agreement with some reports of nanoparticle magnetism
\cite{Chuev2007Generalized-Sto,Lacroix2009Magnetic-hypert}. Therefore, we
first introduced the Stoner-Wohlfarth model with a fixed $H_{\rm c}$
(= 2.5 T) to reproduce the measured hysteresis
loop, however, could not obtain a good fit as shown by a dashed curve in Fig. \ref{hys}(b), because
the experimental curve was broader in the high-field region. Then we
assumed a distribution of $H_{\rm c}$ over a finite range (we assumed the log-normal
distribution) as shown in the inset of Fig. \ref{hys}(b). The calculated
hysteresis loop shown in Fig. \ref{hys}(b) reproduces almost all the characteristics of
the experimental curve, indicating that the SiO$_{2}$-coated FePt
nanoparticles act as non-interacting single-domain
particles whose $H_{\rm c}$ is distributed from $\sim$1 T to $\sim$5
T. The distribution of $H_{\rm c}$ is possibly attributed to the
distribution of structural defects of the FePt
nanoparticles even though high degree of crystallinity is expected to be realized in the SiO$_{2}$-coated FePt nanoparticles as discussed below. Furthermore, because
the Stoner-Wohlfarth model predicts $H_{\rm c}=0.48H_{\rm c(e)}$, where
$H_{\rm c(e)}$ is the $H_{\rm c}$ for the magnetic
field applied along the easy magnetization axis \cite{Stoner1948A-Mechanism-of-}, the $H_{\rm
c(e)}$ of the present FePt nanoparticle sample is estimated as
large as 3.75 T. This value is comparable to that of 4 T for FePt particulate films
\cite{Shima2002Preparation-and} and larger than those of 2.2 T for FePt
(001) dot arrays \cite{Seki2006Dot-size-depend} and 0.75 T for the microfabricated FePt
(001) dots \cite{Wang2008Magnetization-r}, indicating the
large magnetocrystalline anisotropy of the SiO$_{2}$-coated FePt nanoparticles.\\
\indent Finally, we discuss the origin of the observed large $H_{\rm c}$
value for the SiO$_{2}$-coated FePt nanoparticles. The maximum single-domain
size $R_{\rm sd}$ is estimated as $R_{\rm sd}\approx
9\sqrt{\frac{k_{\rm B}T_{\rm C}}{2a_{\rm 0}}K_{\rm 1c}}/\mu_{\rm
0}M_{\rm s}^2$, where $k_{\rm B}$, $T_{\rm C}$, $a_{\rm 0}$, $K_{\rm 1c}$, $\mu_{\rm
 0}$, and $M_{\rm s}$ are the Boltzmann constant, Curie temperature, lattice constant, anisotropy constant, vacuum permeability, and saturation
magnetization, respectively \cite{Coey2010Magnetism-and-M}. Using typical values for FePt nanoparticles of
$K_{\rm 1c}\sim 10^{7}\ {\rm erg/cc} \cite{Usov2012Effective-magne},
 a_{\rm 0}=0.386\ {\rm nm} \cite{Klemmer2002Structural-stud}$
and $M_{\rm s}=2.4\ {\rm \mu_{\rm B}/Fe}$ \cite{Dmitrieva2007Magnetic-moment}, we estimate
the lower limit of $R_{\rm sd}$ as $\sim$100 nm, orders of magnitude
larger than the particle size. Therefore, the SiO$_{2}$-coated FePt
nanoparticles can form in single domains and hence obtain the large
$H_{\rm c}$. If defects are present in the nanoparticles, they hinder
the single-domain formation because the defects act as nucleation
centers \cite{Wang2008Magnetization-r}. Thus, nucleation sites would lead to an enhancement of the $H_{\rm
c}$. In
our fabrication method, the SiO$_{2}$ coating protects the small enough FePt nanoparticle from extra dispersion, deterioration, and oxidization
during the sample preparation, leading to isolated single-domain FePt
particles, and eventually to the large $H_{\rm c}$ value.\\
\indent In conclusion, we have investigated the spin and orbital magnetic moments of Fe
in FePt nanoparticles coated with SiO$_{2}$ using XMCD
measurements. The deduced ratio of the orbital to spin magnetic moments
$M_{\rm orb}/M_{\rm spin}$ = 0.08 is nearly equal to that of FePt
nanoparticles (0.09) condensed from gas phase for which oxidized layers were
removed by {\it in situ} plasmas treatment
\cite{Dmitrieva2007Magnetic-moment}. The magnetization measured by Fe
$L_{\rm 3}$-edge XMCD was saturated around 6 T and $H_{\rm c}$ was as
large as 1.8 T, much larger than the $H_{\rm c}$ ($\sim $0.038 T) of the
gas phase condensed samples
\cite{Dmitrieva2007Magnetic-moment}, and consistent with the result of
the SQUID measurement
\cite{Tamada2007Well-ordered-L1}. The XMCD intensity versus
magnetic field curve was fitted to the Stoner-Wohlfarth model for
non-interacting single-domain nanoparticles whose $H_{\rm c}$ is
distributed between $\sim$1 T to $\sim$5 T.\\
\indent This work was supported by a Grant-in-Aid for Scientific
Research from the JSPS (S22224005) and the ``Quantum Beam Technology
Development Program'' of JST. The experiment was done under the approval
of the Photon Factory Program Advisory Committee (proposal
No. 2010G187), under the Shared Use Program of JAEA Facilities (Proposal
No. 2011A3840), and at JAEA beamline in SPring-8 (proposal No. 2011B3823
and 2012A3823/BL23-SU). SY acknowledges financial support from MEXT
KAKENHI (20104006). YT was supported by JSPS through the Program for
Leading Graduate Schools (MERIT).

%
\end{document}